\documentclass[fleqn,10pt]{wlscirep}
\usepackage[utf8]{inputenc}
\usepackage[T1]{fontenc}
\usepackage{subcaption}
\usepackage{booktabs}
\usepackage{multirow}
\usepackage{multicol}
\usepackage{graphicx}
\usepackage{url,hyperref,lineno,microtype,subcaption}
\title{Modeling Insider Filing Delays in Financial Markets with an Interpretable XGBoost Framework}

\author[1,2,+]{Cheng Huang}
\author[3,+]{Yao Ma}
\author[1]{Fan Gao}
\author[1]{Yutong Liu}
\author[4,5]{Yadi Liu}
\author[6]{Xiaoli Ma}
\author[3]{Ye Aung Moe}
\author[7]{Yuhan Zhang}
\author[2]{Weizheng Xie}
\author[2,9]{Zeyu Han}
\author[1,$\dag$]{Xiangxiang Wang}
\author[1,8]{Hao Wang}
\author[1,$\dag$]{Yongbin Yu}

\affil[1]{University of Electronic Science and Technology of China, Chengdu, 610056, China}
\affil[2]{Southern Methodist University, Dallas, 75205, USA}
\affil[3]{University of Nebraska-Lincoln, Lincoln, 68588, USA}
\affil[4]{Nanyang Technological University, Singapore 639798, Singapore}
\affil[5]{Singapore Management University, Singapore 178899, Singapore}
\affil[6]{Washington State University, Pullman, 99164, USA}
\affil[7]{University at Buffalo, Buffalo, 14260, USA}
\affil[8]{University of Connecticut, Storrs, 06269, USA}
\affil[9]{Georgetown University, Washington, 20057, USA}

\affil[$\dag$]{Corresponding Author: ybyu@uestc.edu.cn}
\affil[+]{these authors contributed equally to this work}

\begin{abstract}

Timely disclosure of insider transactions is a cornerstone of market transparency, yet delays in filing remain widespread and challenging to monitor at scale. This study introduces a comprehensive insider filing delay dataset spanning more than four million Form 4 transactions from 2002 to 2025, enriched with annotations on insider roles, governance attributes, and firm-level indicators. Building on these data, we present a hybrid framework that integrates a state-space encoder with an XGBoost classifier to capture temporal trading patterns while retaining interpretability essential for regulatory auditing. The framework consistently outperforms statistical models, deep sequence learners, and large language model baselines, achieving balanced gains in precision, recall, and F1-score. Feature ablation analyses highlight the predictive importance of insider history, spatiotemporal factors, and governance signals, shedding light on the behavioral drivers of both minor oversights and systematic violations. Beyond accuracy, the dataset and framework establish a reproducible benchmark for studying disclosure compliance, offering regulators and researchers transparent tools to strengthen market integrity.

\end{abstract}

\begin{document}

\maketitle

\section*{Introduction}

Ensuring timely and transparent disclosure of insider trading activities is critical to maintaining fairness and efficiency in financial markets. In the United States, the Securities and Exchange Commission (SEC) mandates that corporate insiders must report their trades through Form 4 filings within two business days of execution \cite{bg-1,finclue,fingpt}. However, strategic delays in disclosure continue to undermine market integrity and investor trust. While prior studies suggest that such delays may be intentional and financially motivated, large-scale empirical analysis remains scarce due to the absence of high-quality, structured datasets with regulatory-grounded labels \cite{bg-2,bg-3}.

To address this gap, we introduce the \textbf{I}nsider \textbf{F}iling \textbf{D}elay (\textbf{IFD}) dataset, the first large-scale, behavior-rich dataset specifically curated for detecting delayed insider filings in violation of SEC rules. IFD contains over 4,051,143 Form 4 filings from 2002 to 2025, annotated with binary labels indicating compliance or violation, and enriched with over 50 features, including insider roles, governance structure, market context, and temporal patterns. This dataset enables the formulation of a new AI task: identifying regulatory disclosure violations as a binary classification problem grounded in financial behavior.

To demonstrate the utility of IFD, we develop \textbf{MaBoost}, a hybrid framework that combines the sequence modeling capabilities of the Mamba \cite{mamba,mamba2} state space model with the interpretability of XGBoost \cite{xgboost}. Our model achieves state-of-the-art performance across multiple datasets, significantly outperforming classical statistical models, deep learning baselines, and large language models (LLMs), while offering transparent, interpretable outputs suitable for regulatory auditing.

This paper contributes a realistic, reproducible foundation for AI-based market surveillance and compliance monitoring, and highlights the importance of behavior-aware modeling in financial regulation. All in all, we make the following key contributions:
\begin{itemize}
    \item We construct IFD, the first and largest publicly available dataset of insider filing behavior, containing over 4,051,143 labeled SEC Form 4 transactions enriched with structured attributes.
    \item We define a novel binary classification task for detecting strategic disclosure violations, grounded in SEC’s two-day reporting regulation.
    \item We propose MaBoost, the hybrid model combining Mamba-based sequence encoding with XGBoost classification, achieving high accuracy and interpretability.
    \item We benchmark IFD across traditional models, deep sequence networks, and LLMs, and show MaBoost consistently outperforms all baselines.
\end{itemize}

\section*{Related Work}

\subsection*{AI for Finance}

Recent advancements in AI for finance have demonstrated promising results in forecasting \cite{rest-1,rest-2,adaint}, anomaly detection \cite{finrl,informer,hirn}, and trading strategy design \cite{grl4fin,finclue,sfinbert}. Classical models like logistic regression \cite{lor}, XGBoost \cite{xgboost}, and random forests \cite{rf} are widely used for structured financial data. Deep learning models, including RNNs \cite{rnn}, LSTMs \cite{lstm}, and Transformers \cite{att}, have further enhanced sequential pattern recognition for price movements \cite{temporal-1,temporal-2}, sentiment analysis \cite{stockm,sentiment}, and fraud detection \cite{gatsf}. Recent works like REST \cite{rest-1,rest-2} and RSR \cite{temporal-2,temporal-1} introduce relational modeling for stock trend forecasting, focusing on cross-stock correlations and event propagation \cite{bg-4}. However, these models primarily aim at market-level predictions and often rely on curated or small-scale datasets. In contrast, our work leverages a real-world, large-scale insider trading dataset, IFD, exceeding one million records, enabling richer modeling of behavioral and regulatory signals beyond market movement.

\subsection*{Report Violations Detection}

Detection of regulatory violations, such as late filings \cite{beyond,finsage} or abnormal trade timing \cite{mcs,gegan}, remains underexplored in comparison to mainstream financial prediction tasks. Prior studies have treated such violations as outlier detection problems or framed them within rule-based or heuristic systems with limited generalizability \cite{deep,fingpt}. While some supervised learning efforts exist, they typically lack sufficient labeled data and fail to capture temporal or contextual nuances in insider behavior \cite{bg-1,adamw}. To address these limitations, we propose a novel framework, MaBoost, that not only incorporates temporal and relational structure across trading sequences but also benefits from a uniquely labeled dataset covering a broad spectrum of violation cases \cite{fingpt}. MaBoost outperforms traditional baselines and aligns with the goals of compliance automation and intelligent regulatory monitoring.

\section*{Methodology}

MaBoost is designed as a modular, hybrid architecture that combines sequential pattern extraction and structured classification for the detection of insider filing violations. The framework consists of two main components: a Mamba-based sequence encoder and an XGBoost classifier.

Given a historical transaction sequence $\mathbf{X}_i$, the Mamba sequence encoder processes each time step via a linear state-space mechanism:

\begin{equation}
\mathbf{h}_t = \mathbf{A}_t \mathbf{h}_{t-1} + \mathbf{B}_t \mathbf{x}_i^{(t)}, \quad \mathbf{z}_t = \mathbf{C}_t \mathbf{h}_t
\end{equation}
The sequence of outputs $\{ \mathbf{z}_1, \ldots, \mathbf{z}_T \}$ is aggregated into a fixed-length representation $\mathbf{h}_i$, either by mean pooling or using the final state:

\begin{equation}
\mathbf{h}_i = \text{Aggregate}(\{ \mathbf{z}_1, \ldots, \mathbf{z}_T \})
\end{equation}
The vector $\mathbf{h}_i$ is then passed to an XGBoost classifier composed of $M$ regression trees:

\begin{equation}
\hat{y}_i = f(\mathbf{h}_i) = \sum_{m=1}^{M} f_m(\mathbf{h}_i)
\end{equation}
The predicted probability is obtained via sigmoid activation:

\begin{equation}
\hat{p}_i = \sigma(\hat{y}_i)
\end{equation}
and the final label is determined as:

\begin{equation}
\hat{y}_i^{\text{label}} = \mathbb{I}(\hat{p}_i \geq \tau)
\end{equation}
The model is trained by minimizing a regularized logistic loss:

\begin{equation}
\mathcal{L} = \sum_{i=1}^{n} l(y_i, \hat{p}_i) + \sum_{m=1}^{M} \Omega(f_m)
\end{equation}

where $l(\cdot)$ is the binary cross-entropy loss and $\Omega(f_m)$ is the complexity regularization term. This formulation allows MaBoost to capture temporal behavioral patterns while retaining interpretable, tree-based decision logic.

As shown in Table~\ref{notation}, all the terms used in this section are summarized below.

\begin{table}[ht]
\begin{center}
\caption{Key Mathematical Notations in MaBoost}
\label{notation}
\begin{tabular}{l|l}
\toprule
\textbf{Symbol} & \textbf{Description} \\
\midrule
$\mathbf{A}_t$ & State transition matrix in the Mamba encoder \\
$\mathbf{B}_t$ & Input projection matrix in the Mamba encoder \\
$\mathbf{C}_t$ & Output projection matrix in the Mamba encoder \\
$\mathbf{x}_i^{(t)}$ & Feature vector at time $t$ for insider $i$ \\
$\mathbf{X}_i$ & Historical transaction sequence of insider $i$ \\
$\mathbf{h}_t$ & Hidden state at time step $t$ in the Mamba encoder \\
$\mathbf{z}_t$ & Output vector at time $t$ from the Mamba encoder \\
$\mathbf{h}_i$ & Aggregated behavioral representation of insider $i$ \\
$f$ & Overall XGBoost classifier function \\
$f_m$ & The $m$th decision tree in the XGBoost ensemble \\
$\hat{y}_i$ & Raw prediction score (logit) for sample $i$ \\
$\hat{p}_i$ & Predicted probability after sigmoid activation \\
$\hat{y}_i^{\text{label}}$ & Final binary classification label \\
$y_i$ & Ground-truth label \\
$\mathcal{L}$ & Overall loss function \\
$l(y_i, \hat{p}_i)$ & Binary cross-entropy loss for sample $i$ \\
$\Omega(f_m)$ & Regularization penalty for the $m$th tree \\
$\sigma(\cdot)$ & Sigmoid activation function \\
$\mathbb{I}(\cdot)$ & Indicator function for thresholding \\
$\tau$ & Classification threshold (typically 0.5) \\
\bottomrule
\end{tabular}
\end{center}

\end{table}

\section*{Dataset and Implementation}

\subsection*{Dataset}
We construct our insider trading dataset by extracting stock transaction records from WRDS\footnote{https://wrds-www.wharton.upenn.edu/} and processing them in SAS\footnote{www.sas.com}. Following established standards, we retain trades with valid cleanse codes (R, H, C, L, I), exclude amended filings and option-related transactions, and focus on open market purchases and sales (TRANCODE = P or S) \cite{bg-1}. Filings with transaction dates after the report date are removed. To measure reporting delays, we merge the data with an LLM-generated SEC business calendar (2002–2025), incorporating a Python-based two-day filing deadline. Unlike prior work, we preserve trade-level granularity without insider-day aggregation to support machine learning tasks.

\subsubsection*{IFD} \textbf{I}nsider \textbf{F}iling \textbf{D}elay contains 4,051,143 transactions, involving over 7,633 firms and 15,573 insiders, covering open market transactions disclosed via Form 4 filings by corporate insiders in U.S. public companies from 2002 to 2025. Among these trades, roughly 17.4\% (21,482 transactions) are classified as filing violations, meaning the disclosure occurred after the SEC’s legally mandated deadline, a window reduced to two business days after the Sarbanes-Oxley Act (SOX) in 2002. Filing violations are further categorized based on the length of the delay and the historical behavior of the insider. About 77\% of the violations are considered oversight violations, typically delayed by three or fewer business days and committed by insiders who do not frequently violate filing requirements. In contrast, 23\% of the violations are deemed intentional violations, characterized by longer delays ($\geq$4 business days) and committed by insiders who violate at least 95\% of the time.

Also, as shown in Table~\ref{Stat-Rates}, we also looked at different internal roles within the company and their violation rates. Classified statistics on the job titles of employees who violated the rules can help provide a clearer understanding of the approximate proportions when analyzing the motives behind these violations later. This is crucial in determining whether the violation falls under the category of 'negligent violation' or 'intentional violation. 

\begin{table}[ht]
\centering
\caption{Insider Roles and Violation Rates}
\label{Stat-Rates}
\begin{tabular}{l|ccc}
\toprule
\textbf{Role} & \textbf{Total Trades} & \textbf{Violation Rate} & \textbf{Violations} \\ 
\midrule
CEO & 12,843 & 10.88\% & 1,397 \\
Corporate Suite & 21,415 & 10.63\% & 2,277 \\
Beneficial Owners & 18,402 & 22.24\% & 4,093 \\
Other Insiders & 8,472 & 16.59\% & 1,405 \\ 
\bottomrule
\end{tabular}
\end{table}

One sample of IFD has a total of 52 attributes as shown in Table~\ref{detail-ifd}. These fields encompass identity information, transaction details, form types, data verification, delays, and adjustments necessary for SEC insider trading reports. Finance experts of our team thoroughly review and check all data, guaranteeing its authenticity and accuracy. The IFD dataset includes six attribute categories: (1) Basic Identification and Date Information, (2) Company and Security Information, (3) Form and Transaction Details, (4) Transaction Codes and Data Cleansing, (5) Timing and Delay Information, and (6) Additional Fields.

\begin{table}[ht]
\centering
\tiny
\caption{Details of One Sample of IFD}
\label{detail-ifd}
\scalebox{1.12}{
\begin{tabular}{l|l}
\toprule
\textbf{Attribute} & \textbf{Value} \\
\midrule
DCN & 02720220 \\
TRANDATE & 08/01/2002 \\
SEQNUM & 1 \\
PERSONID & 16078646 \\
Owner & SHAVIN HELENE B \\
Rolecode1 & O \\
Rolecode2 & VP \\
Rolecode3 & C \\
Rolecode4 & - \\
Address1 & 1 ROCKEFELLER PLAZA \\
Address2 & STE 1430 \\
City & NEW YORK \\
State & NY \\
Zipcode & 10020 \\
Country & - \\
Phone & - \\
Cname & HARRIS \& HARRIS GROUP INC \\
Cnum & H202680000 \\
CUSIP6 & 413833 \\
CUSIP2 & 10 \\
Ticker & TINY \\
SECID & 4537 \\
Sector & 01 \\
Industry & 02 \\
FORMTYPE & 4 \\
Acqdisp & A \\
Optionsell & - \\
Ownership & D \\
Sharesheld & 500.0000 \\
Sharesheld\_Adj & 500.0000 \\
Shares & 500.0000 \\
Shares\_Adj & 500.0000 \\
TPRICE & 2.79 \\
Tprice\_Adj & 2.7900 \\
TRANCODE & P \\
Sectitle & COM \\
Amend & - \\
Cleanse & R \\
FDATE & 05/29/2012 \\
CDATE & 08/07/2002 \\
MAINTDATE & 05/27/2012 \\
SECDATE & 08/06/2002 \\
SIGDATE & 08/06/2002 \\
TRANDATE\_AR & 08/01/2002 \\
ACQDISP\_AR & A \\
TPRICE\_AR & 2.79 \\
TRANCODE\_AR & P \\
Gap\_Days & -3589 \\
SEC\_Business\_Day & 2002-08-05 \\
SEC\_Business\_Day\_Lag2 & 2002-08-01 \\
Delay & 1 \\
\midrule
Id & 0 \\
\bottomrule
\end{tabular}}
\end{table}

Basic Identification and Date Information includes DCN (record ID), TRANDATE (transaction date), SEQNUM (record sequence), PERSONID (insider ID), OWNER (insider name), ROLECODE1–4 (insider roles), ADDRESS1/2, CITY, STATE, ZIPCODE, COUNTRY (location details), and PHONE. Company and Security Information covers CNAME (company name), CNUM (internal ID), CUSIP6 and CUSIP2 (issue identifiers), TICKER (stock symbol), SECID (security ID), and classification codes for SECTOR and INDUSTRY. Form and Transaction Details include FORMTYPE (Form 3, 4, 5, or 144), ACQDISP (acquisition 'A' or disposition 'D'), OPTIONSELL (option-related sale flag), and OWNERSHIP (beneficial ownership: Direct or Indirect). SHARES and SHARES\_ADJ record transaction volume (raw and adjusted), while SHARESHELD and SHARESHELD\_ADJ indicate post-trade holdings. TPRICE and TPRICE\_ADJ reflect the transaction price per share (raw and adjusted). Transaction Codes and Data Cleansing contain TRANCODE / TRANCODE\_AR (transaction type, e.g., ‘P’ for purchase), SECTITLE (security type), AMEND (amendment flag), and CLEANSE (data confidence level). Timing and Delay Information includes FDATE, CDATE, MAINTDATE (file/create/maintenance dates), SECDATE (SEC receipt date), SIGDATE (signature date), and TRANDATE\_AR / ACQDISP\_AR / TPRICE\_AR (as-reported values). gap\_days measures transaction-to-filing delay. SEC\_Business\_Day and SEC\_Business\_Day\_Lag2 mark filing timeliness, while delay quantifies reporting lag in days. Additional Fields include the Rule 10b5-1 flag (plan-based trade), Market Value of Transaction (Form 144 estimate), and Proposed Number of Shares (expected to be sold under Form 144). Id is the internal unique identifier for the record, which is the label (1 means illegal operation). For the Table~\ref{detail-ifd},  - means none.

All in all, IFD can be used to analyze insider trading behavior, filing violations, market reactions, corporate governance, regulatory effectiveness, and firm performance, providing valuable insights into the dynamics of insider compliance and its impact on financial markets. Other important information can be found in Supplementary Material.

\subsection*{Implementation}

\subsubsection*{Hyperparameter}
Table~\ref{hyper} summarizes the hyperparameter configurations used for training the MaBoost framework, which integrates the Mamba state space encoder with an XGBoost classifier. For the Mamba module, we adopt a lightweight sequence encoder with bidirectional modeling and GELU activation \cite{gelu}. For XGBoost, we apply standard configurations suitable for imbalanced binary classification tasks, with regularization and subsampling strategies. This configuration balances model capacity, interpretability, and computational efficiency, and performs robustly across all evaluation scenarios. The training epoch is 200 and the validation is done by ten-fold cross validation.

\begin{table}[ht]
\centering
\caption{Hyperparameter Settings for MaBoost}
\label{hyper}
\begin{tabular}{l|l|l|l}
\toprule
\multicolumn{2}{c|}{\textbf{Mamba}} & \multicolumn{2}{c}{\textbf{XGBoost}} \\
\midrule
Perparameter & Value & Perparameter & Value \\
\midrule
\texttt{d\_model} & \texttt{256} & \texttt{objective} & \texttt{binary\_logistic} \\
\texttt{n\_layers} & \texttt{4} & \texttt{tree\_method} & \texttt{hist} \\
\texttt{ssm\_rank} & \texttt{4} & \texttt{max\_depth} & \texttt{10} \\
\texttt{dropout} & \texttt{0.1} & \texttt{min\_child\_weight} & \texttt{10} \\
\texttt{activation} & \texttt{GELU} & \texttt{gamma} & \texttt{0.8} \\
\texttt{prenorm} & \texttt{True} & \texttt{subsample} & \texttt{0.8} \\
\texttt{use\_bidirectional} & \texttt{True} & \texttt{colsample\_bytree} & \texttt{0.8} \\
\texttt{seq\_len} & \texttt{100} & \texttt{learning\_rate} & \texttt{0.1} \\
\texttt{bias} & \texttt{True} & \texttt{n\_estimators} & \texttt{3000} \\
\texttt{norm\_epsilon} & \texttt{0.00005} & \texttt{reg\_alpha} & \texttt{0.1} \\
\texttt{tuning\_strategy} & \texttt{Optuna} & \texttt{reg\_lambda} & \texttt{1.0} \\
\bottomrule
\end{tabular}
\end{table}

\subsubsection*{Hardware} The hardware specifications for training and testing include 4 4090-Ti GPUs (4 $\times$ 24GB), 64GB of RAM, 8 CPU cores per node, and a total of 6 nodes. 

\section*{Experimental Result}

\subsection*{Experimental Setup}

For the IFD dataset, input features can be configured in three modes: (1) Equal Weight; (2) Constraint Condition; (3) Suspected Violation. In the \textit{Equal Weight} setting, all 52 attributes are treated uniformly, providing a baseline to evaluate the model’s capacity for pattern extraction without prior assumptions. The \textit{Constraint Condition} setting incorporates empirical insights by downplaying feature groups with limited standalone impact (e.g., \textit{Spatiotemporal}~\cite{bg-1}). MaBoost's adaptive weighting compensates by capturing their interactions with stronger signals. The \textit{Suspected Violation} setting focuses on subtle behavioral cues such as trading history or delayed filings that gain importance when aligned with abnormal financial patterns. MaBoost’s co-attentive design effectively amplifies such weak but correlated indicators, leading to substantial improvements in precision and recall. This underscores the importance of retaining diverse features for reliable detection of nuanced violations \cite{bg-3}. 

We evaluate experimental results using Recall, Precision, and F1-Score, which are standard metrics for binary classification tasks ($\times$100\%). We highlight the best-performing value in bold and the second-best in underline. The training epoch is 500 and the validation is done by ten-fold cross validation.

\subsubsection*{Baseline Model}

We choose these following methods or models:  Linear Regression \cite{lir}, Logistic Regression \cite{lor}, Seq2Seq \cite{seq2seq}, RNN  \cite{rnn}, Bi-RNN \cite{birnn}, LSTM \cite{lstm}, Bi-LSTM \cite{bilstm}, GRU \cite{gru}, Transformer \cite{att}, Mamba \cite{mamba}, Decision Tree \cite{dt}, Random Forest \cite{rf}, XGBoost \cite{xgboost}, BERT\cite{bert,tispell}, CINO \cite{cino} and RoBERTa \cite{roberta,fmsd}. We also conducted comparative experiments using a set of encoder-decoder frameworks, including combinations of CNN \cite{cnn} (ResNet-152 \cite{resnet}), Vision Transformer (ViT) \cite{vit}, and Convolutional Vision Transformer (ConViT) \cite{convit} encoders with various decoders (e.g., RNN \cite{rnn}, LSTM \cite{lstm}, Transformer \cite{att} and Mamba \cite{mamba}). For parameter optimization, we selected tool Optuna\footnote{https://optuna.org/} as the optimization method. 

\begin{table*}[!ht]
\centering 
\tiny
\caption{Hyperparameters Setting of LLMs} 
\label{hyper-1} 
\scalebox{1.1}{
\begin{tabular}{l|l|ccc|ccc|cccccccc} 
\toprule 
\multirow{2.5}{*}{\textbf{LLM}} & \multirow{2.5}{*}{\textbf{Version}} & \multicolumn{3}{c|}{\textbf{Prompt-based N-shot}} & \multicolumn{3}{c|}{\textbf{Embedding-based Classification}} &  \multicolumn{8}{c}{\textbf{Fine-tuning with the Classification Head}}\\
\cmidrule{3-16} & & \textbf{Temp} & \textbf{Top\_p} & \textbf{Stream} & \textbf{Dim Change}  & \textbf{Method} & \textbf{Classifier} & \textbf{Optimizer} & \textbf{LR} & \textbf{DR} & \textbf{GC} & \textbf{HS}  & \textbf{BS} & \textbf{MSL} & \textbf{Epoch} \\
\midrule
\multirow{4}{*}{GPT}  & 3.5-Turbo  & 1.0 & 1.0 & False & 1536$\rightarrow$1024 & TC           & MLP & AdamW & 10$^{-5}$ & 0.1 & 1.0 & 256  & 8 & 512 & 5 \\
  & 4O & 1.0 & 1.0 &  False & 1536$\rightarrow$1024 & TC          & MLP & AdamW & 10$^{-5}$ &  0.1 & 1.0  & 256 & 8 & 512 & 5  \\
  & O1-mini & 1.0 & 1.0 &  True & 1536$\rightarrow$1024 & TC          & MLP & AdamW & 10$^{-5}$ &  0.1 & 1.0 &  256 & 8 & 512 & 5 \\
  & O1 & 1.0 & 1.0 &  True & 1536$\rightarrow$1024 & TC          & MLP & AdamW & 10$^{-5}$ & 0.1 & 1.0 &  256 & 8 & 512 & 5 \\
\midrule
\multirow{3}{*}{LlaMA}  & 3.1-405B & 0.6 & 0.9 &  False & 4096$\rightarrow$1024 & PCA     & MLP & AdamW & 10$^{-5}$ & 0.1 & 1.0 &  256 & 8 & 512 & 5  \\
   & 3.1-70B & 0.6 & 0.9 &  False & 4096$\rightarrow$1024 & PCA    & MLP & AdamW & 10$^{-5}$ & 0.1 & 1.0 & 256  & 8 & 512 & 5 \\
   & 3.1-8B  & 0.6 & 0.9 &  False & 4096$\rightarrow$1024 & PCA    & MLP & AdamW & 10$^{-5}$ & 0.1 & 1.0 & 256  & 8 & 512 & 5  \\
\midrule
\multirow{3}{*}{Qwen}  & 2.5-72b & 0.7 & 0.8 &  False & 4096$\rightarrow$1024 & LP          & MLP & AdamW & 10$^{-5}$ & 0.1 & 1.0 & 256  & 8 & 512 & 5  \\
  & 2.5-32b & 0.7 & 0.8 &  False & 4096$\rightarrow$1024 & LP         & MLP & AdamW & 10$^{-5}$ & 0.1  & 1.0 & 256  & 8 & 512 & 5 \\
  & 2.5-7b & 0.7 & 0.8 &  False & 4096$\rightarrow$1024 & LP         & MLP & AdamW & 10$^{-5}$ & 0.1 & 1.0 & 256  & 8 & 512 & 5 \\
\midrule
\multirow{2}{*}{DeepSeek}  & R1 & 1.0 & None &  True & 4096$\rightarrow$1024 & MP+PCA        & MLP & AdamW & 10$^{-5}$ & 0.1 & 1.0 & 256  & 8 & 512 & 5 \\
  & V3 & 1.0 & None &  False & 4096$\rightarrow$1024 & MP+PCA       & MLP & AdamW & 10$^{-5}$ & 0.1 & 1.0 & 256  & 8 & 512 & 5 \\
\midrule
Claude & 3-5-Sonnet & 1.0  & None & False & 1024 & -               & MLP & AdamW & 10$^{-5}$ & 0.1 & 1.0 & 256  & 8 & 512 & 5 \\
Gemini  & 1.5-Flash & None & 0.95 & False & 3072$\rightarrow$1024 & TC  1024         & MLP & AdamW & 10$^{-5}$ & 0.1 & 1.0 & 256  & 8 & 512 & 5 \\
\bottomrule 
\end{tabular}}
\end{table*}

\subsubsection*{Large Language Model}

We also evaluate several LLMs, categorized as open-source and closed-source. Open-source models include the LLaMA \cite{llama-1,llama-2,llama-3}, Qwen \cite{qwen,qwen-2,qwen-2.5,qwen-3}, and DeepSeek \cite{v3,r1} families, while closed-source models include GPT \cite{gpt3,gpt4o}, Claude \cite{claude3}, and Gemini \cite{gemini}. For open-source models, we evaluate the Qwen-2.5 (2.5-7B, 2.5-32B, 2.5-72B) and DeepSeek families via their official APIs, and the LLaMA (3.1-8B, 3.1-70B, 3.1-405B) family through the LLaMA-API platform. Closed-source models are similarly evaluated through their respective official APIs. These evaluation details will be fully documented in the final version to ensure transparency and reproducibility.

We evaluate the IFD dataset using three classification paradigms based on LLMs: (1) Prompt-based Zero/Few-shot Inference; (2) Embedding-based Classification; (3) Fine-tuning with the Classification Head.

Table~\ref{hyper-1} presents the hyperparameters for prompt-based zero-shot and few-shot classification. In this setting, LLMs are guided by natural language prompts to generate classification labels without additional training. For embedding-based classification, LLMs act as frozen feature extractors, producing contextual embeddings from hidden states. These embeddings are then passed to an external binary classifier, ensuring efficiency and flexibility across tasks. To ensure fair comparison, all embeddings are projected to 1024 dimensions using PCA~\cite{pca-1,pca-2}, linear projection~\cite{pca-1,att}, mean pooling~\cite{mp-1,mp-2}, and truncation~\cite{att,gpt4o}. The classifier is a two-layer MLP~\cite{mlp}. In the fine-tuning setup, a classification head (also an MLP~\cite{mlp}) is trained jointly with the LLM for binary classification, enabling task-specific representation learning. This approach typically achieves higher accuracy but at the cost of increased computation and potential overfitting. Cross-entropy loss~\cite{cross} is used, with cosine decay and 10\% warm-up, and optimization is performed using AdamW~\cite{adamw}.

\noindent\textbf{Note:} Temperature = Temp; Linear Projection = LP; Mean Pooling = MP; Truncate = TC; Learning Rate = LR; Dropout Rate = DR; Gradient Clipping = GC; Hidden Size = HS; Batch Size = BS; Max Sequence Length = MSL.

\subsection*{Comparative Experiment}

\subsubsection*{Baseline Model}

Table~\ref{ce-os} presents the performance comparison of various models across three input configurations on the IFD dataset. Among all models, MaBoost consistently achieves the best performance across all three input types, significantly outperforming baselines in terms of F1-Score. Notably, MaBoost demonstrates strong stability, with Recall, Precision, and F1-Score remaining closely aligned and exhibiting no significant fluctuations, indicating balanced decision boundaries and robust generalization. Transformer-based hybrids (e.g., ConViT-Transformer and ConViT-Mamba) also show competitive results, particularly under the \textit{Constraint Condition} and \textit{Suspected Violation} settings. Traditional models such as Logistic Regression and XGBoost perform reasonably well but fall short in capturing complex interaction patterns. 

\begin{table*}[ht]
\centering
\caption{Comparative Experiment Among Models on IFD }
\label{ce-os}
\scalebox{0.85}{
\begin{tabular}{l|ccc|ccc|ccc}
\toprule
\multirow{2.5}{*}{\textbf{Model}}  &  \multicolumn{3}{c|}{\textbf{Equal Weight}}  & \multicolumn{3}{c|}{\textbf{Constraint Condition}} & \multicolumn{3}{c}{\textbf{Suspected Violation}} \\
\cmidrule{2-10}& \textbf{Precision}  & \textbf{Recall}  &  \textbf{F1-Score} & \textbf{Precision}  & \textbf{Recall}  &  \textbf{F1-Score} &  \textbf{Precision}  & \textbf{Recall}  &  \textbf{F1-Score} \\
\midrule
Linear Regression  &  52.39 & 12.54 &  20.24 & 76.15  & 14.42 & 24.25 & 67.79 & 13.72  & 22.75\\
Logistic Regression  & 59.89  & 34.24 &  42.18  & 85.60 & 57.07  & 60.79 & 69.72 & 41.75  & 45.98\\
Seq2Seq & 76.67  & 74.68 &  77.02 & 93.97 & 94.85 & 93.15 & 92.38 & 93.14  & 92.89 \\
RNN  & 77.01  & 77.56 &  78.07 & 90.36 & 94.27 &  91.51 & 78.98 & 79.97  & 81.24 \\
Bi-RNN  &  81.19 & 84.27 &  83.39 & 93.97 & 94.85 &  93.15 & 90.27 & 91.69  & 90.89\\
LSTM  & 78.58  & 79.42 &  79.23 & 90.24 & 94.28 &  91.50 & 84.87 & 86.65  & 85.93 \\
Bi-LSTM  &  83.37 & 85.13 &  84.29 & 95.05 & 99.73 &  97.33 & 91.28 & 92.97  & 92.09 \\
GRU  &  82.19 & 83.27 &  83.09 & 88.88 & 94.28 &  89.15 & 84.19 & 86.07  & 85.74\\
Transformer  &  87.32 & 87.98 &  87.07 & 96.71 & \underline{99.81} &  98.29 & 91.17 & 68.94  & 71.28\\
Mamba  & 86.98  & 87.27 &  87.15 & 98.23 & 98.74 & \underline{98.69} & \underline{93.59} & 94.27  & 93.89\\
Decision Tree  & 77.86  & 54.98 &  61.21 & 96.66 & 44.08 &  60.55 & 81.92 & 83.57  & 82.62 \\
Random Forest  & 79.86  & 82.98 &  83.27 & 93.17 & 94.60 &  92.88 & 86.17 & 80.93  & 76.84 \\
XGBoost  & 88.71  & 90.03 &  90.17 & 97.91 & 97.87 &  97.65 & 91.21 & 90.17  & 87.28 \\
BERT  & 91.37  & \underline{92.44} &  \underline{91.89}  & 97.98 & 97.97 & 97.78 & 92.31 & 92.92  & 91.23 \\
CINO &  90.34 & 91.09 &  90.87 & 97.25 & 84.85 &  90.63 & 88.97 &  89.91 & 89.37 \\
RoBERTa  & \underline{92.32}  & 91.12 &  90.89  & 97.03  & 78.97 & 87.07 & 93.05 &  \underline{94.75} & \underline{93.98} \\
\midrule
CNN-Seq2Seq & 81.23  & 82.17 &  81.98 & 96.66 & 44.08 &  60.65 & 92.38 & 89.17  & 86.18 \\
CNN-RNN  & 82.47  & 83.18 &  83.05 & 95.29 & 66.76 &  73.79 & 91.42 &  72.39 & 79.69 \\
CNN-LSTM  & 81.39  & 82.73 &  82.07 & 95.98 & 95.95 &  94.23 & 77.65 & 56.42  & 62.27\\
CNN-GRU  & 82.13  & 84.38 &  82.45 & 96.69 & 71.99 &  79.42 & 85.91 & 87.53  & 86.29 \\
ViT-RNN  & 84.49  & 83.18 &  71.49 & 97.30 & 75.99 &  85.34 & 87.27 &  72.72 & 78.26 \\
ViT-LSTM  &  86.53 & 89.41 &  87.87 & 97.88 & 89.41 &  93.18 & 88.91 & 92.57 & 90.14\\
ViT-Transformer  & 89.96  & 92.35 &  91.84 & 97.98 & 97.99 &  \underline{97.81} & 89.98 & 47.91  & 58.95\\
ViT-Mamba  &  87.44 & 89.71 &  90.37 & 98.24 & 93.72 & 95.85 & 90.37 &  83.62 & 84.13 \\
ConViT- RNN  & 87.98  & 89.31 &  88.79 & 95.94 & 98.95 & 97.20 & 91.07 & 92.16  & 91.79 \\
ConViT-LSTM  & 89.03  & 91.27 &  90.76 & 96.36 & 65.09 &  71.99 & 92.78 & 78.94  & 82.16\\
ConViT-Transformer  &  \underline{93.16} & \underline{93.98} &  92.07 & 94.28 & \underline{98.97} & 97.05  & 93.47 & \underline{95.19}  & \underline{94.87}\\
ConViT-Mamba  &  92.28 & 93.34 &  \underline{92.89} & \underline{98.64} & 91.32 & 94.36 & \underline{94.16} &  89.79 & 90.39\\
\midrule
MaBoost  & \textbf{94.93}  & \textbf{95.48} &  \textbf{94.79} & \textbf{99.09} & \textbf{99.85} & \textbf{99.47} & \textbf{96.24} &  \textbf{97.08} & \textbf{96.76} \\
\bottomrule
\end{tabular}}
\end{table*}

\subsubsection*{Large Language Model}

\begin{figure}[htbp]
    \centering
    \begin{subfigure}[b]{0.45\textwidth}
        \includegraphics[width=\linewidth]{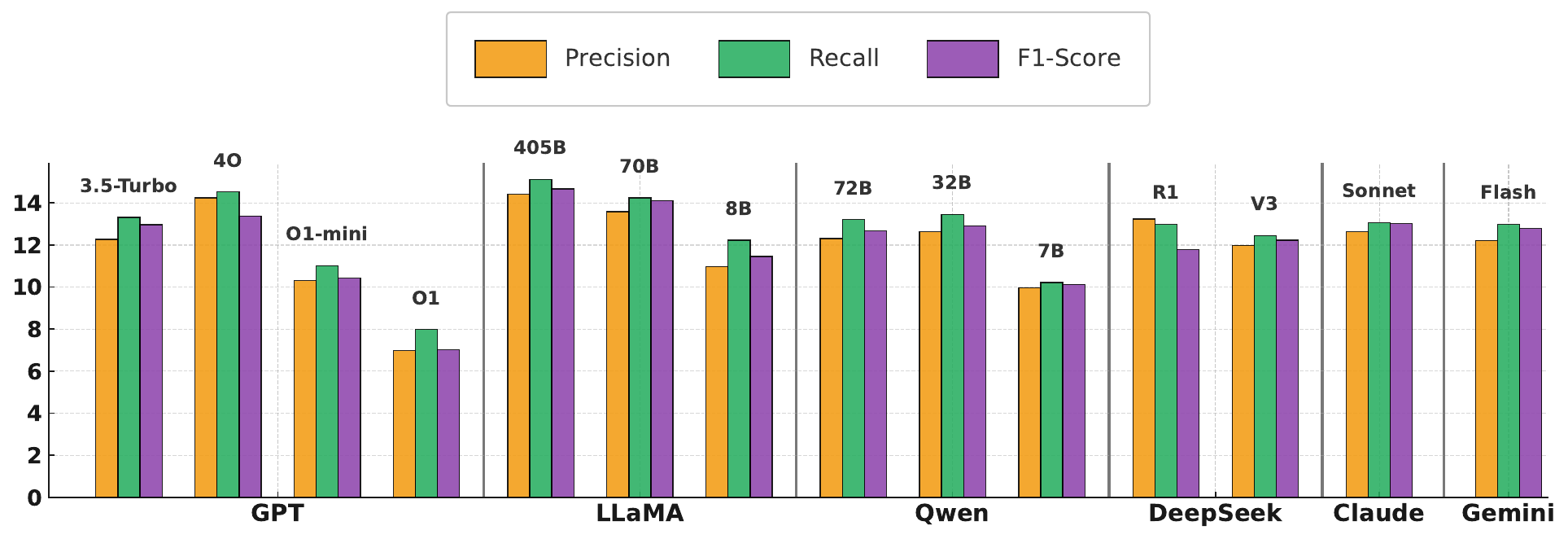}
        \caption{Zero-shot}
        \label{zero}
    \end{subfigure}
    \hfill
    \begin{subfigure}[b]{0.45\textwidth}
        \includegraphics[width=\linewidth]{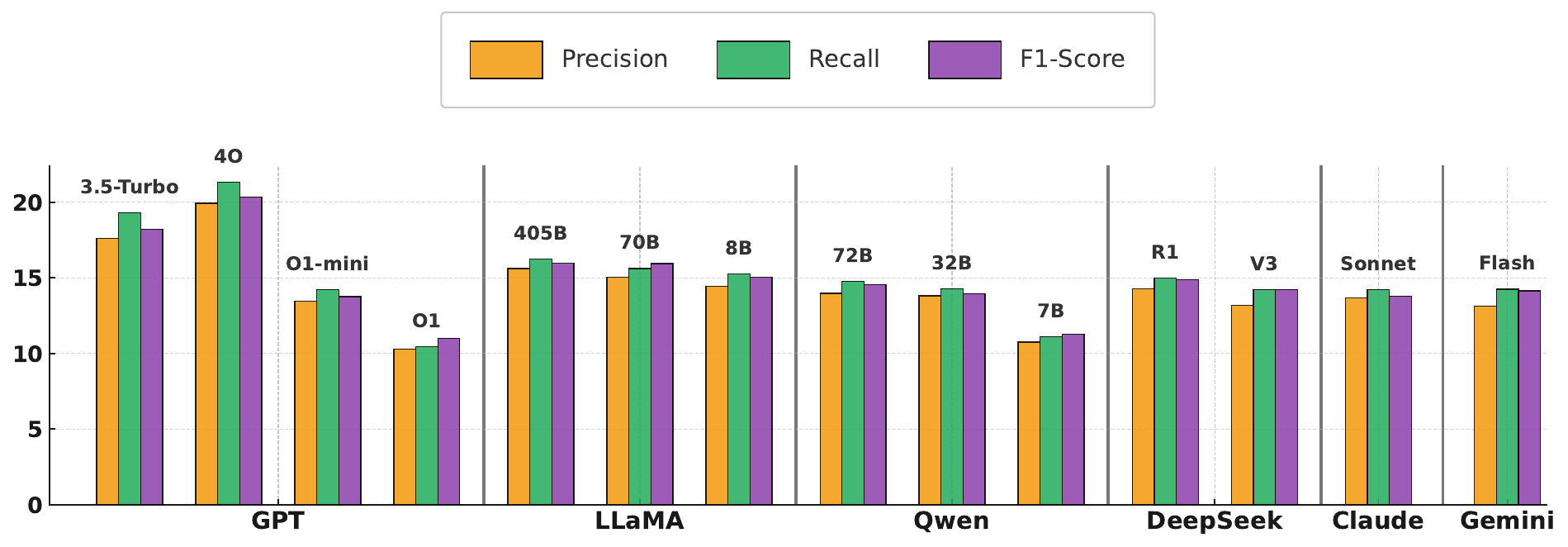}
        \caption{One-shot}
        \label{one}
    \end{subfigure}

    \begin{subfigure}[b]{0.45\textwidth}
        \includegraphics[width=\linewidth]{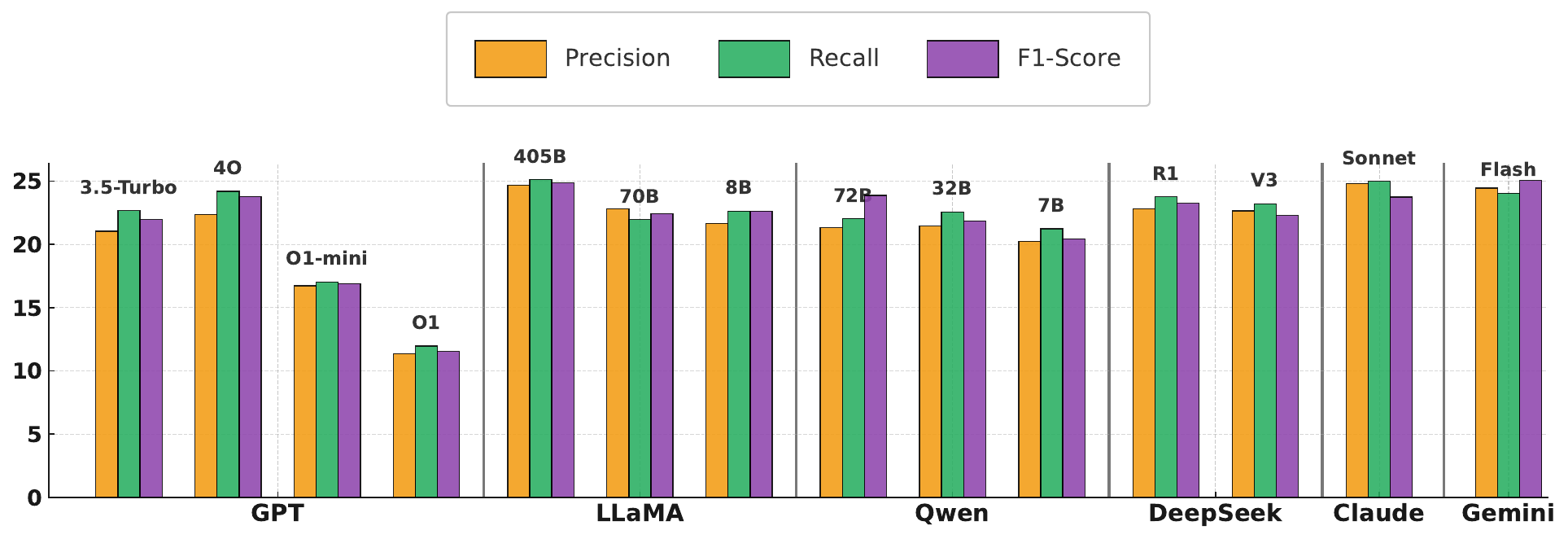}
        \caption{Two-shot}
        \label{two}
    \end{subfigure}
    \hfill
    \begin{subfigure}[b]{0.45\textwidth}
        \includegraphics[width=\linewidth]{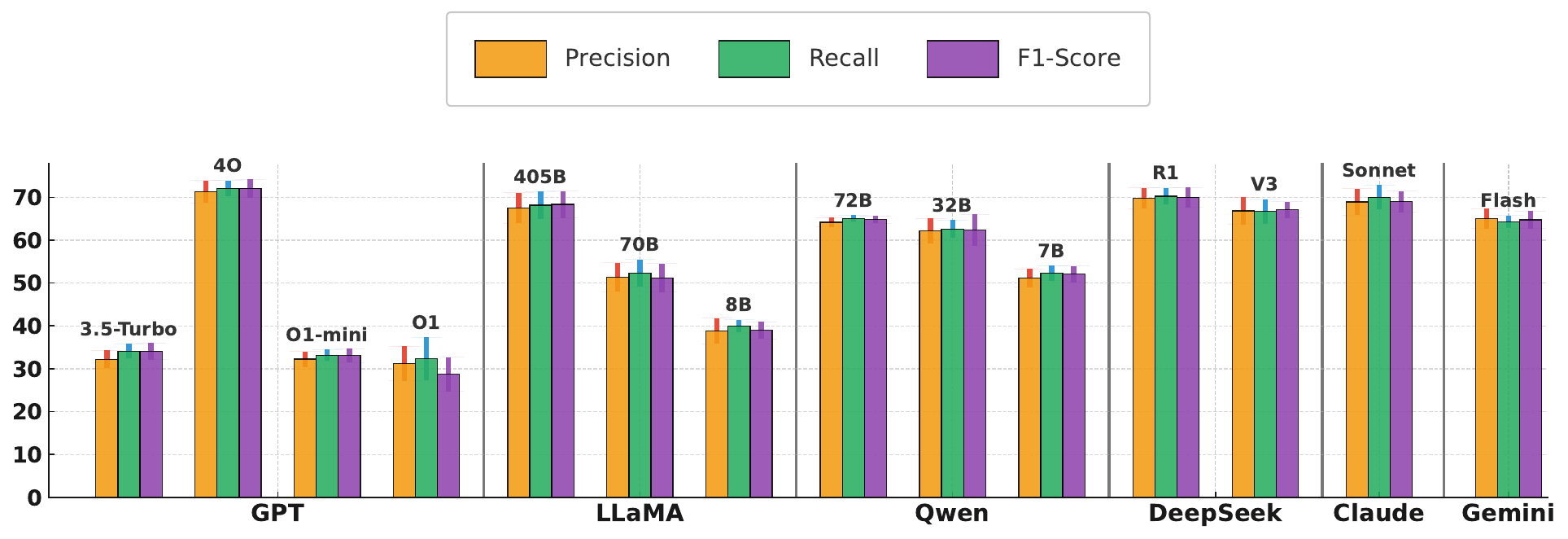}
        \caption{N-shot (Best)}
        \label{best}
    \end{subfigure}

    \caption{Prompt-based N-shot Experiment of LLMs on IFD}
    \label{N}
\end{figure}

As shown in Figure~\ref{N}, although LLMs have shown remarkable capabilities in natural language understanding, their performance on structured financial datasets like IFD remains significantly limited. This is primarily due to three factors: (1) LLMs are not optimized for tabular data reasoning, which involves numerical patterns and relational attributes rather than textual semantics; (2) they are prone to overfitting when the task lacks sufficient labeled context for in-context learning; and (3) zero-shot and few-shot prompting underperform due to the semantic gap between instruction-style prompts and highly structured features. This observation underscores the necessity of tailored AI models in compliance-critical financial domains, rather than relying on general-purpose LLMs.

\begin{figure}[htbp]
    \centering
    \begin{subfigure}[b]{0.32\textwidth}
        \includegraphics[width=0.95\linewidth]{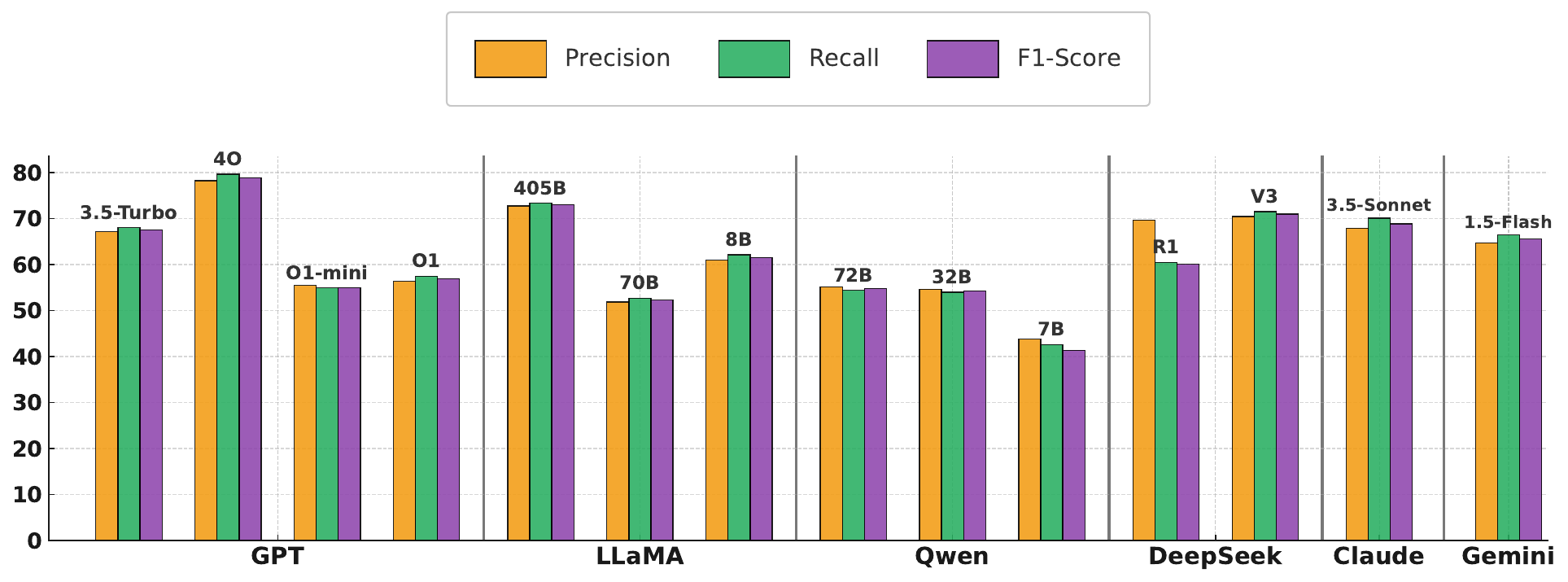}
        \caption{Equal Weight}
        \label{EW}
    \end{subfigure}
    \hfill
    \begin{subfigure}[b]{0.32\textwidth}
        \includegraphics[width=0.95\linewidth]{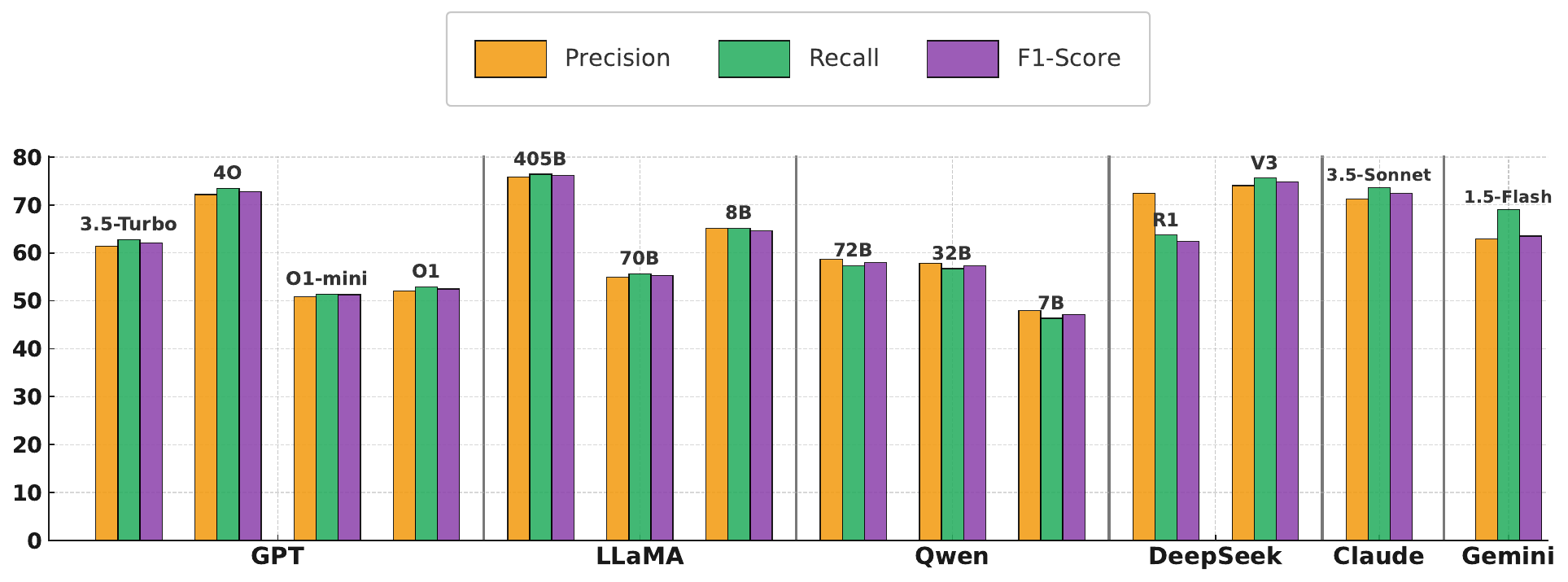}
        \caption{Constraint Condition}
        \label{CC}
    \end{subfigure}
    \hfill
    \begin{subfigure}[b]{0.32\textwidth}
        \includegraphics[width=0.95\linewidth]{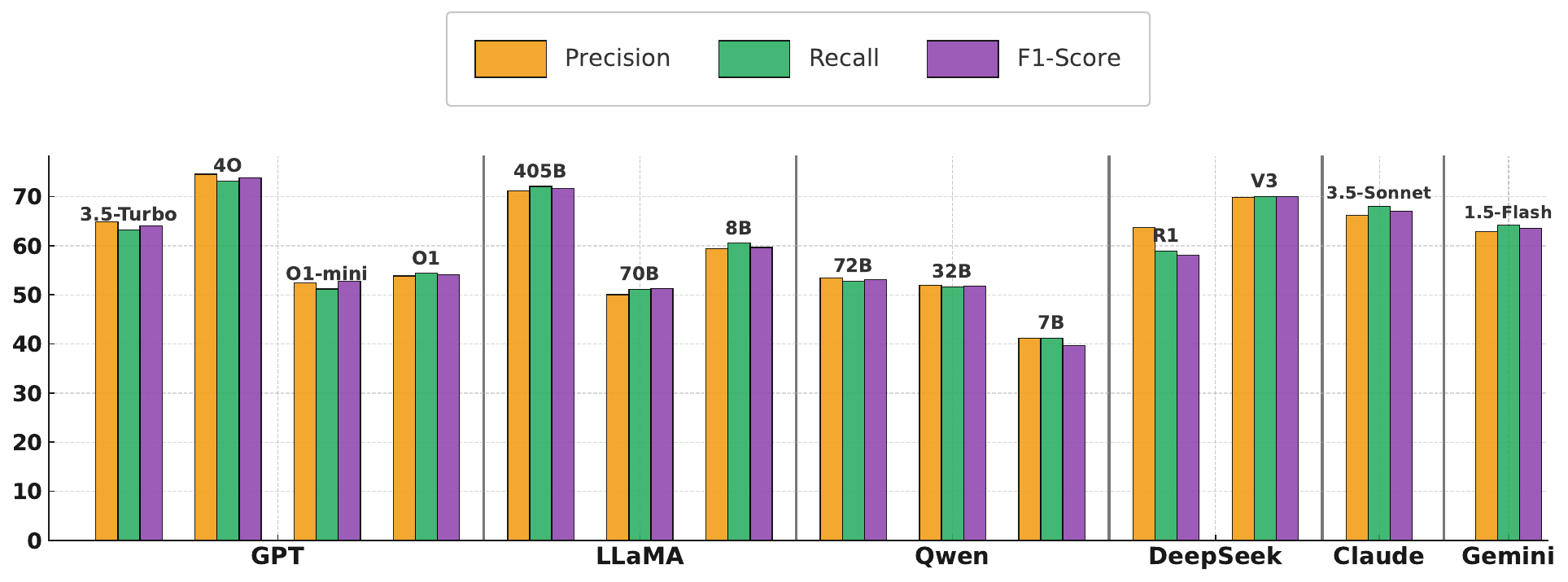}
        \caption{Suspected Violation}
        \label{SV}
    \end{subfigure}

    \caption{Embedding-based Classification of LLMs on IFD}
    \label{embed}
\end{figure}

Figure~\ref{embed} presents the embedding-based classification performance of various LLMs on the IFD. Among all LLMs, GPT-4O and LLaMA-3.1-405B consistently achieve strong results, with GPT-4O obtaining the highest F1-Score under Equal Weight and Suspected Violation tasks. In contrast, smaller-scale models such as Qwen-2.5-7b and GPT-O1-mini exhibit relatively lower performance across all settings. The inclusion of constraint-based logic notably improves most models’ precision, recall, and F1 scores, highlighting the benefits of incorporating domain-specific knowledge into the classification process. 

\begin{figure}[htbp]
    \centering
    \begin{subfigure}[b]{0.32\textwidth}
        \includegraphics[width=0.95\linewidth]{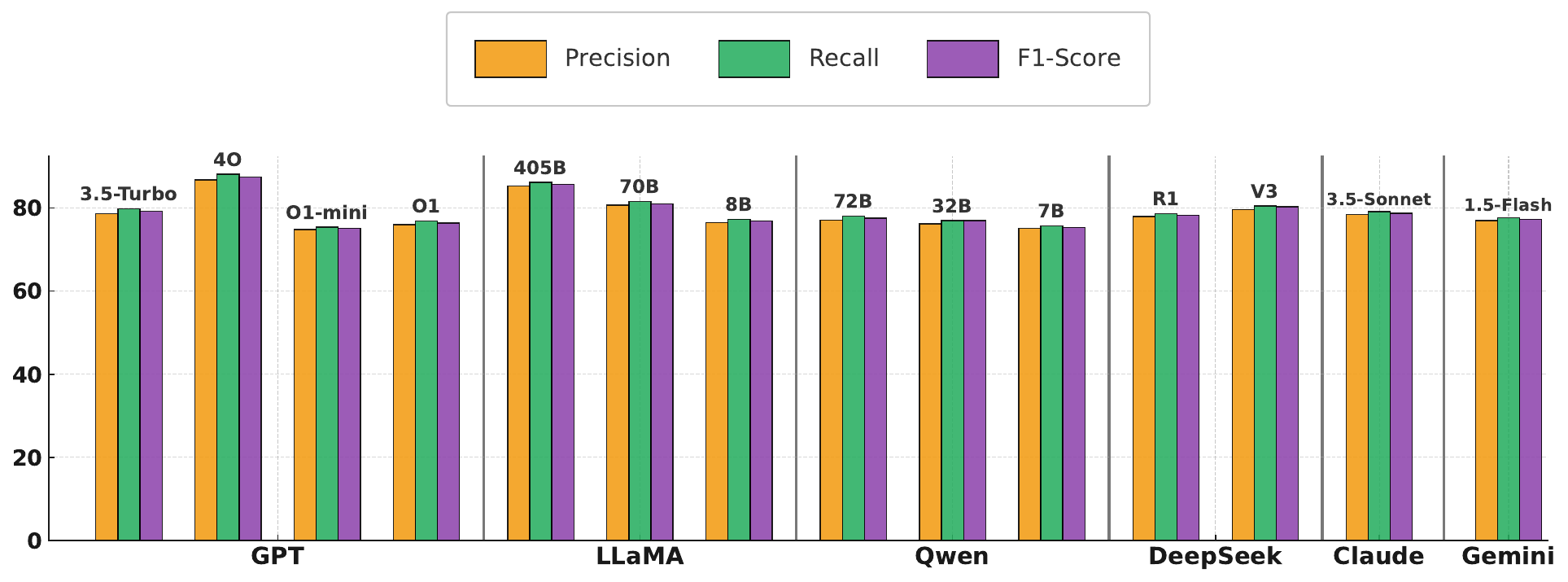}
        \caption{Equal Weight}
        \label{EW-1}
    \end{subfigure}
    \hfill
    \begin{subfigure}[b]{0.32\textwidth}
        \includegraphics[width=0.95\linewidth]{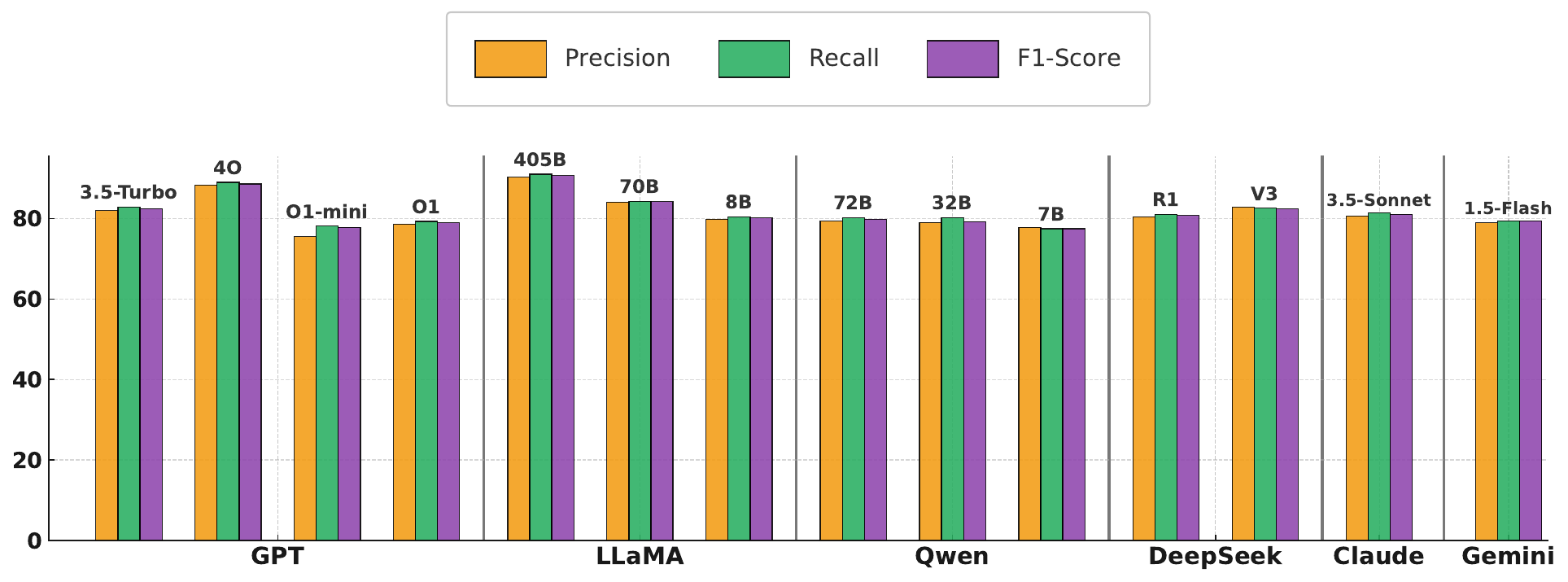}
        \caption{Constraint Condition}
        \label{CC-2}
    \end{subfigure}
    \hfill
    \begin{subfigure}[b]{0.32\textwidth}
        \includegraphics[width=0.95\linewidth]{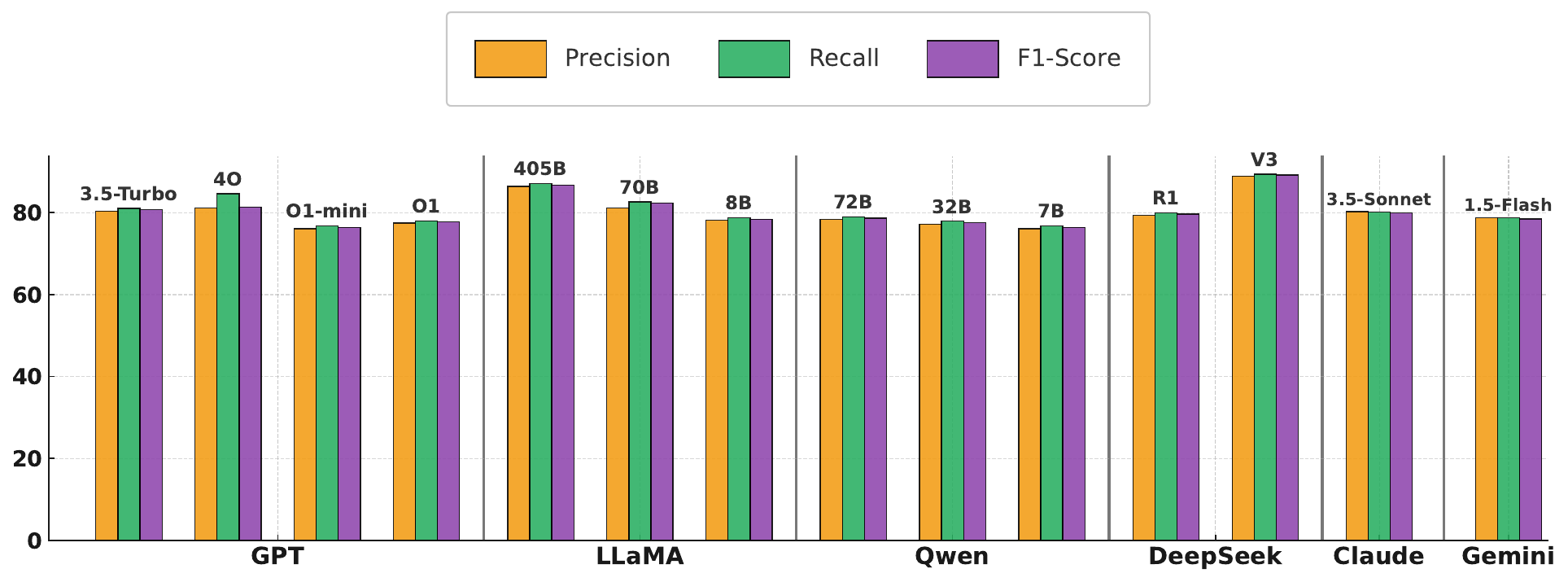}
        \caption{Suspected Violation}
        \label{SV-3}
    \end{subfigure}

    \caption{Full Fine-tuning Performance of LLMs on IFD}
    \label{futn}
\end{figure}

Figure~\ref{futn} reports the full fine-tuning results of various LLMs on the IFD dataset using a unified MLP classification head. GPT-4O and LLaMA-3 405B consistently outperform others, with GPT-4O achieving the highest F1-score (87.44) under Equal Weight and LLaMA-3 405B reaching 90.67 under Constraint Condition. DeepSeek-V3 shows notable strength in Suspected Violation detection (F1: 89.13), while smaller models such as GPT-3.5, Qwen-2.5-7B, and LLaMA-3.1-8B perform comparatively lower. Claude and Gemini maintain stable performance.

Compared with LLMs, MaBoost outperforms all across all settings, demonstrating its superiority in regulatory classification tasks and highlighting the strong discriminatory capability of the IFD dataset, enabling comprehensive evaluation and comparison among state-of-the-art models.
\begin{table*}[!ht]
\centering
\caption{Ablation Experiment on IFD Feature Groups with MaBoost}
\label{tab:ifd_ablation}
\scalebox{0.67}{
\begin{tabular}{l|ccc|ccc|ccc}
\toprule
\multirow{2.5}{*}{\textbf{Feature Removed}} &  \multicolumn{3}{c|}{\textbf{Equal Weight}} & \multicolumn{3}{c|}{\textbf{Constraint Condition}} &  \multicolumn{3}{c}{\textbf{Suspected Violation}} \\
\cmidrule{2-10}& \textbf{Precision } & \textbf{Recall } & \textbf{F1-Score} & \textbf{Precision } & \textbf{Recall } & \textbf{F1-Score} & \textbf{Precision } & \textbf{Recall } & \textbf{F1-Score} \\
\midrule
Insider History (InsiderRatio, FirmRatio) & 88.14 & 89.57 & 86.09 & 95.38 & 96.82 & 95.79 & 91.37 & 92.41 & 91.76 \\
Trade Characteristics (TradeValue, Delay) & 91.02 & 91.76 & 91.33 & 96.57 & 90.35 & 96.92 & 82.65 & 84.07 & 83.21 \\
Governance (BlockholderRatio, HHI) & 92.53 & 92.62 & 92.11 & 87.22 & 87.91 & 87.52 & 73.18 & 75.01 & 74.06 \\
Financial Health (ROA, Leverage, Tobin's Q) & 93.71 & 93.85 & 93.71 & 98.36 & 99.08 & 98.65 & 84.81 & 85.72 & 85.09 \\
Spatiotemporal (Ln(Distance), Gap Days) & 92.89 & 92.41 & 82.33 & 87.74 & 88.46 & 77.98 & 64.16 & 65.47 & 64.79 \\
\midrule
Full Dataset &  \textbf{94.93}  & \textbf{95.48} &  \textbf{94.79} & \textbf{99.09} & \textbf{99.85} & \textbf{99.47} & \textbf{96.24} &  \textbf{97.08} & \textbf{96.76} \\
\bottomrule
\end{tabular}}
\end{table*}

\begin{table*}[!ht]
\centering
\caption{Ablation Experiment of MaBoost }
\label{ce-ab}
\scalebox{0.85}{
\begin{tabular}{l|ccc|ccc|ccc}
\toprule
\multirow{2.5}{*}{\textbf{Model}}  &    \multicolumn{3}{c|}{\textbf{Equal Weight}} & \multicolumn{3}{c|}{\textbf{Constraint Condition}} &  \multicolumn{3}{c}{\textbf{Suspected Violation}}\\
\cmidrule{2-10}& \textbf{Precision}  & \textbf{Recall}  &  \textbf{F1-Score} & \textbf{Precision}  & \textbf{Recall}  &  \textbf{F1-Score} & \textbf{Precision}  & \textbf{Recall}  &  \textbf{F1-Score}\\
\midrule
Decision Tree  & 77.86  & 54.98 &  61.21 & 96.66 & 44.08 &  60.55 & 81.92 & 83.57  & 82.62 \\
+ Transformer  & 81.74  & 62.37 &  68.64 & 97.34 & 48.12 &  63.67 & 83.68 & 85.33  & 84.49 \\
+ Mamba        & 84.09  & 66.24 &  72.72 & 97.89 & 50.71 &  66.12 & 85.94 & 87.85  & 86.73 \\
\midrule
Random Forest  & 79.86  & 82.98 &  83.27 & 93.17 & 94.60 &  92.88 & 86.17 & 80.93  & 76.84 \\
+ Transformer  & 83.92  & 85.33 &  84.60 & 96.23 & 96.94 &  96.32 & 87.88 & 84.21  & 80.52 \\
+ Mamba        & 86.30  & 87.95 &  87.01 & 97.02 & 97.76 &  97.23 & 90.23 & 86.15  & 82.79 \\
\midrule
XGBoost  & 88.71  & 90.03 &  90.17 & 97.91 & 97.87 &  97.65 & 91.21 & 90.17  & 87.28 \\
 + Transformer & 91.24 & 92.35 & 91.79 & 98.16 & 98.92 & 98.51 & 93.14 & 94.01 & 93.52 \\
Mamba  & 86.98  & 87.27 &  87.15 & 98.23 & 98.74 & 98.69 & 93.59 & 94.27  & 93.89\\
+ NoConv   & 84.33 & 84.87 & 84.60 & 97.03 & 97.64 & 97.31 & 91.45 & 92.03 & 91.71 \\
+ NoRes    & 83.27 & 83.91 & 83.58 & 96.82 & 96.93 & 96.73 & 90.83 & 91.27 & 91.04 \\
+ d8       & 85.42 & 85.63 & 85.49 & 97.36 & 97.88 & 97.55 & 92.03 & 92.86 & 92.44 \\
+ dt1      & 85.76 & 86.07 & 85.92 & 97.68 & 98.04 & 97.85 & 92.28 & 93.04 & 92.66 \\
+ GELU  & 85.91 & 86.38 & 86.14 & 97.52 & 98.02 & 97.77 & 92.74 & 93.55 & 93.14 \\
\midrule
MaBoost  &  \textbf{94.93}  & \textbf{95.48} &  \textbf{94.79} & \textbf{99.09} & \textbf{99.85} & \textbf{99.47} & \textbf{96.24} &  \textbf{97.08} & \textbf{96.76} \\
\bottomrule
\end{tabular}}
\end{table*}

\subsection*{Ablation Experiment}
\subsubsection*{IFD} To assess the value of each feature group in MaBoost, we conduct an ablation study by removing individual categories, as detailed in Table~\ref{tab:ifd_ablation}. Performance drops with the exclusion of any group, confirming their complementary roles. Notably, removing \textbf{Insider History} or \textbf{Spatiotemporal} features under the \textit{Suspected Violation} setting leads to a marked decrease in F1-score, underscoring their central importance for violation detection. In contrast, \textit{Financial Health} and \textit{Governance} features are more influential in the \textit{Equal Weight} and \textit{Constraint Condition} scenarios. These results highlight the necessity of a comprehensive feature set for robust violation identification.

\subsubsection*{MaBoost}

To assess the impact of different modules and design choices in MaBoost, we conduct ablation studies on three base classifiers, Decision Tree, Random Forest, and XGBoost, progressively augmenting them with Transformer and Mamba backbones, as shown in Table~\ref{ce-ab}. Adding a Transformer consistently improves decision boundary learning, while replacing it with Mamba yields further gains, reflecting Mamba’s superior ability to model long-range dependencies in tabular data. We further analyze the Mamba backbone through controlled variants. Key Mamba components include state dimension, time interpolation, convolutional input projection, gating and normalization strategies, and activation function. Removing the convolutional projection or residual connections significantly reduces performance, underscoring their roles in spatial encoding and stable optimization. Varying the state dimension or time step resolution provides marginal improvements, while switching the activation from \texttt{SiLU} to \texttt{GELU} slightly degrades results, indicating better alignment of Mamba’s default activation with its architecture.  Collectively, these ablations confirm that each architectural element is essential to Mamba’s effectiveness, driving both the robustness and generalization of MaBoost in complex rule-based classification tasks.

\section*{Discussion}

One potential concern regarding the reported results is that the classification performance, with F1-scores exceeding 99\%, might appear unusually high for real-world financial tasks. This outcome is not primarily a reflection of model overfitting, but rather of the structural clarity of the regulatory rule and the informative nature of the dataset. Insider filing delays are defined by a strict SEC disclosure requirement (two business days), which yields a well-bounded binary outcome. In this context, features such as insider identity, trading history, and governance attributes directly capture behavioral and organizational patterns strongly associated with compliance outcomes. 

Moreover, the proposed framework integrates a state-space encoder to represent temporal patterns and XGBoost to provide interpretable decision boundaries. While the model achieves very high scores on cross-validation and hold-out sets, its transparency allows auditors and researchers to confirm that predictions are grounded in meaningful variables rather than spurious correlations. Importantly, even small margins of improvement are significant for regulators, since delayed disclosures can undermine market integrity and investor confidence. 

We emphasize that the contribution of this study is not only the high predictive accuracy but also the release of a large-scale, regulatory-grade dataset and an interpretable modeling framework. These resources enable reproducibility, comparative benchmarking, and further exploration of subtle forms of non-compliance that may not be captured by simpler models. Future work can extend this approach to international markets and multi-modal filings, where the compliance landscape is more complex and classification performance is unlikely to saturate at similar levels.

\section*{Conclusion}

This work contributes a regulatory-grade dataset and an interpretable modeling framework for the systematic study of insider filing delays. By linking structured disclosure attributes with an XGBoost-based classifier, the proposed approach achieves both predictive accuracy and transparency, aligning with the needs of regulatory oversight. The results underscore that disclosure violations, though governed by clear rules, reveal distinct behavioral and organizational patterns that can be effectively modeled with hybrid methods. Importantly, the release of the dataset and benchmarks enables reproducibility and comparative evaluation, offering a foundation for future research. While the current framework addresses U.S. markets, extending this approach to international contexts, multi-modal filings, and causal inference represents a promising direction for advancing the detection and understanding of financial disclosure compliance.

\section*{Future Work}
In future work, we aim to (1) expand the IFD dataset to cover international jurisdictions and multilingual filings; (2) incorporate unstructured textual disclosures (e.g., Form 10-K narratives) using document-level embeddings; and (3) explore causal learning frameworks to uncover deeper compliance logic beyond statistical correlation, facilitating proactive regulatory interventions.

\section*{Acknowledgments}
This work was supported in part by the National Natural Science Foundation of China under Grants 62276055 and 62406062,in part by the Sichuan Science and Technology Program under Grant 2023YFG0288, in part by the Natural Science Foundation of Sichuan Province under Grant 2024NSFSC1476,in part by the National Science and Technology Major Project under Grant 2022ZD0116100,in part by the Sichuan Provincial Major Science and Technology Project under Grant 2024ZDZX0012.

\section*{Data Availability Statement}
All data and codes are available: \url{https://github.com/CH-YellowOrange/MaBoost-and-IFD}.

\bibliography{sample}

\end{document}